\documentclass[12pt]{article}
\usepackage{graphicx}
\usepackage{subfigure}
\usepackage{amsfonts}
\usepackage{amssymb,amsmath}

\newcommand{\bea}{\begin{eqnarray}}
\newcommand{\eea}{\end{eqnarray}}

\makeatletter
\@addtoreset{equation}{section}

\makeatother

\usepackage{multicol}
\usepackage[usenames,dvipsnames]{xcolor}
\definecolor{rosso}{cmyk}{0,1,1,0.3}
\definecolor{verde}{cmyk}{0.8,0,0.6,0.25}
\definecolor{bluc}{cmyk}{1,0.4,0,0.1}
\definecolor{blucc}{cmyk}{0.8,0.3,0,0}

\def\be{\begin{equation}}
\def\ee{\end{equation}}

\def\({\left(}
\def\){\right)}
\def\1{^{(1)}}
\def\2{^{(2)}}

\def\<{\langle}
\def\>{\rangle}

\begin{document}

\begin{titlepage}

\begin{flushright}
UT-12-45\\
\end{flushright}

\vskip 3cm

\begin{center}

{\Large \bf 
Dissipative Effects on Reheating after Inflation
}

\vskip .6in

{
Kyohei Mukaida$^{(a)}$
and
Kazunori Nakayama$^{(a,b)}$
}

\vskip .3in

{\em
$^a$Department of Physics, University of Tokyo, Bunkyo-ku, Tokyo 113-0033, Japan \vspace{0.2cm}\\
$^b$Kavli Institute for the Physics and Mathematics of the Universe,
University of Tokyo, Kashiwa 277-8583, Japan \\
}

\end{center}

\vskip .5in

\begin{abstract}

The inflaton must convert its energy into radiation after inflation, which, in a conventional scenario,
is caused by the perturbative inflaton decay.
This reheating process would be much more complicated in some cases:
the decay products obtain masses from an oscillating inflaton and thermal environment, and 
hence the conventional reheating scenario can be modified.
We study in detail processes of particle production from the inflaton, their subsequent thermalization
and evolution of inflaton/plasma system by taking dissipation of the inflaton in a hot plasma into account. 
It is shown that the reheating temperature is significantly affected by these effects.

\vskip .45in

\end{abstract}

\end{titlepage}

\setcounter{page}{1}

\section{Introduction} 
\label{sec:introduction}

The idea of inflation~\cite{Guth:1980zm,Linde:1981mu} has now become a part of standard cosmological evolution scenario.
It provides beautiful explanations for the nearly flat isotropic/homogeneous Universe and
the origin of primordial density fluctuation, which results in rich observed cosmological structures.
For successful inflation, the energy of the inflaton, which drives the inflationary expansion of the Universe,
must be transferred to the radiation consisting of hot standard model (SM) plasma.
This process, called {\it reheating}, is rather an unknown aspect of inflation~\cite{Allahverdi:2010xz},
partly because the process is model dependent and partly because the era of reheating is difficult to be explored
observationally.
However, the reheating temperature $T_{\rm R}$, corresponding to the temperature of the hot plasma
at the beginning of the radiation dominated era, is an important characteristics of inflation model since it often determines
the efficiency of leptogenesis/baryogenesis and the abundance of (unwanted) relics such as the gravitino and moduli.

In a conventional picture, the inflaton is assumed to have a coupling to light fields and perturbatively decays into them.
Produced light SM particles are thermalized and constitute radiation component of the Universe thereafter.
In this case, the reheating temperature simply depends on the perturbative decay rate of the inflaton:
\begin{equation}
	T_{\rm R}^{\rm (w.b.)} \equiv \left( \frac{90}{\pi^2 g_*} \right)^{1/4}\sqrt{ \Gamma_\phi^0 M_{\rm pl}},   \label{TR_wb}
\end{equation}
where $g_*$ is the relativistic degrees of freedom at the temperature $T=T_{\rm R}^{\rm (w.b.)}$,
$\Gamma_\phi^0$ denotes the inflaton decay rate evaluated at the vacuum and $M_{\rm pl}$ is the reduced Planck scale.
Here we defined this quantity, $T_{\rm R}^{\rm (w.b.)}$, as ``would-be-reheating temperature''.

However, this simple picture does not hold for some inflation models for the following reasons.
First, the inflaton is oscillating around its potential minimum,
and hence masses of coupled particles also oscillate with time, which would invalidate the use of the inflaton decay rate
at the vacuum~\cite{Kofman:1994rk,Shtanov:1994ce,Felder:1998vq}.
Second, before the complete decay of the inflaton, the Universe is often already filled with high-temperature plasma.
Thus light particles, including SM particles, obtain thermal masses and should be treated as quasi-particles,
which would significantly modify the inflaton decay rate into these particles.
On this second aspect, it was pointed out that large thermal masses of decay products prevent
the inflaton decay and the temperature of plasma cannot be as high as the inflaton mass (divided by a coupling constant)~\cite{Kolb:2003ke}.
This is not true: in high-temperature environment, the quasi-particles obtain thermal widths
and the inflaton dissipates its energy into thermal plasma
as was explicitly shown in Refs.~\cite{Yokoyama:2005dv,Drewes:2010pf} in the context of reheating.
Intuitively this is understood as a result of efficient scattering processes between inflaton 
and quasi-particles in thermal plasma.
Thus actual thermal history would be much more complicated
and, in particular, the reheating temperature would be significantly different from the estimate (\ref{TR_wb}).

In this paper we address the issues of thermalization and reheating after inflation in detail.
We start from the inflaton oscillation just after inflation and show how particle production and their thermalization occur.
Then we study the evolution of the inflaton oscillation and plasma by taking into account the inflaton dissipation
in high-temperature plasma and also the non-perturbative particle production,
until the inflaton dissipates all its energy, after which the radiation dominated Universe begins.
Formulations for these effects are found in our previous paper~\cite{Mukaida:2012qn},
where dynamics of scalar fields in thermal environment was studied in detail.

It is shown that deviation from the conventional reheating scenario becomes more prominent for smaller inflaton mass,
and hence we mainly focus on low-scale inflation model.
Such low scale inflation is actually realized in the Higgs inflation~\cite{Bezrukov:2007ep,GarciaBellido:2008ab}, 
where the SM Higgs field plays a role of inflaton, since its mass around the vacuum is weak scale.
Some supersymmetric (SUSY) inflation models are also classified into this category:
{\it e.g.}, MSSM inflation~\cite{Allahverdi:2006iq}, alchemical inflation~\cite{Nakayama:2012gh} or others.

In Sec.~\ref{sec:reheating}, we briefly discuss particle production and their thermalization.
Dissipation coefficients in thermal plasma are also listed.
Using these ingredients, we study the evolution of inflaton and plasma system, and determine the
reheating temperature in Sec.~\ref{sec:reh_after_inf}.
We conclude in Sec.~\ref{sec:conclusion}.

\section{Particle production and dissipation} 
\label{sec:reheating}

\subsection{Setup} 

Let us consider a following simple setup where 
the inflaton $\phi$ interacts with light fields $\chi$ via Yukawa interaction:
\begin{align}
	{\cal L} = {\cal L}_{\rm kin} 
	 - \frac{1}{2} m_\phi^2 \phi^2 + \lambda \phi \( \bar \chi_{\rm L} \chi_{\rm R} + {\rm h.c.} \)
	 + {\cal L}_{\rm other}
	 \label{Lag}
\end{align}
where $\lambda$ is a coupling constant taken to be real and positive, ${\cal L}_{\rm kin}$ denotes canonical
kinetic terms, and ${\cal L}_{\rm other}$ denotes the other light degrees of freedom
including gauge bosons. The bare mass of $\chi$ is neglected in what follows.
We also assume that the $\chi$ fields are charged under some gauge groups
and they interact with other light degrees of freedom via these gauge interactions.
The coupling constant $\lambda$ and gauge coupling $g$ are assumed to
be smaller than unity.\footnote{
	It is possible to consider the case where Yukawa interactions
	dominantly connect the $\chi$ fields with the other light degrees of freedom.
	The following calculation does not change much if the coupling constant $g$ is reinterpreted as the Yukawa coupling.
}
We also define $\alpha\equiv g^2/(4\pi)$ for later convenience.

Note that the model (\ref{Lag}) should be regarded as a representative model which correctly describes essential features 
of more general class of models. It is straightforward to extend the model as
\begin{equation}
	{\cal L} = {\cal L}_{\rm kin} 
	 - \sum_{k,l}\frac{1}{2} m_{\phi_{kl}}^2 \phi_{kl}^2 
	 +\sum_{k,l} \lambda_{kl} \phi_{kl} \( \bar \chi_{{\rm L},k} \chi_{{\rm R},l} + {\rm h.c.} \)
	 + {\cal L}_{\rm other},
\end{equation}
where integers $k,l$ include possible flavor and gauge indices.
In particular, the inflaton can have gauge charges.
If one of the scalar fields $\phi_{kl}$ obtains a large field value and takes a role of inflaton,
the dynamics of inflaton is effectively described by a simplified model (\ref{Lag}).
In a SUSY model, there should be a inflaton coupling to bosons 
as ${\cal L} = \lambda^2|\phi|^2|\tilde\chi|^2$ with $\tilde\chi$ denoting the scalar partner of $\chi$.
Inclusion of this coupling does not modify the following arguments,
as long as we restrict ourselves to the case where the parametric resonant phenomena do not occur (see Sec.~\ref{sec:inst_prht}).

The initial amplitude of inflaton at the end of inflation, $\phi_i$, is taken as a free parameter,
without specifying the inflaton potential beyond $\phi_i$.
We only assume that the subsequent coherent oscillation can be well
described by the quadratic potential: $m_\phi^2 \phi^2 /2$.
As mentioned in the Introduction, for the small amplitude case, {\it i.e.} $m_\phi \gg \lambda \tilde \phi$
with $\tilde\phi$ representing the amplitude of oscillating inflaton field,\footnote{
	$\tilde \phi$ and $\phi (t)$ represent the amplitude and oscillating field value respectively.
}
the effect of thermal plasma on the inflaton dissipation was already studied by~\cite{Yokoyama:2005dv,Drewes:2010pf}. 
Hence, hereafter, we mainly concentrate on
the large amplitude case: $m_\phi \ll \lambda \tilde \phi$.
The purpose of this paper is to clarify the thermalization and reheating process in this class of models,
especially the reheating temperature, by solving the evolution of inflaton and plasma system.

The basic ingredients that we will use in the following discussion
are found in detail in our previous paper~\cite{Mukaida:2012qn} and here we briefly repeat the results.
Let us see what happens after inflation in the following subsections.

\subsection{Particle production and thermalization} 
\label{sec:inst_prht}

\subsubsection{Instant preheating} 

After the inflation, the inflaton starts to oscillate around its potential
minimum with an initial amplitude $\phi_i$.
Hence, the coupled field $\chi$ has an amplitude-dependent dispersion relation:
$\omega_\chi^2 = k^2 + m^\chi_{\rm th}(T)^2 + \lambda^2 \phi^2(t)$
where $m^\chi_{\rm th}(T)$ denotes a possible thermal mass.
Initially the thermal mass vanishes ($m^\chi_{\rm th} = 0$), 
since there is no background plasma.
If $\lambda\phi_i \ll m_\phi$, the perturbative decay of the inflaton into $\chi$ creates thermal plasma as in a conventional scenario.
On the other hand, if $\lambda\phi_i \gg m_\phi$, such a process is kinematically blocked due to the $\phi$-dependent mass of $\chi$
at the most time domains in each one oscillation of the inflaton.
Instead, the following non-perturbative particle production process becomes important.\footnote{
	The perturbative inflaton decay into gauge bosons through one-loop process involving $\chi$ field is possible,
	but its efficiency is lower than that of the non-perturbative particle production.
}

The efficient non-perturbative particle production occurs when
the adiabaticity of the coupled $\chi$ fields is broken down~\cite{Kofman:1994rk}:
$|\dot \omega_\chi / \omega_\chi^2 | \gg 1$. From this inequality,
the following condition is obtained:
\begin{align}
	\lambda \tilde \phi \gg {\rm max} \left[ m_\phi, \frac{g^2 T^2}{m_\phi} \right]. 
	\label{eq:np_cond}
\end{align}
Here $\tilde \phi$ stands for the amplitude of $\phi$.
Importantly, Eq.~\eqref{eq:np_cond} implies that if the thermal mass 
$m^\chi_{\rm th}(T) \sim gT$ becomes as large as $k_\ast \equiv (\lambda \tilde \phi m_\phi )^{1/2}$,
then the non-perturbative particle production does not occur~\cite{Mukaida:2012qn}.
If Eq.~\eqref{eq:np_cond} is met, the $\chi$'s modes below 
the typical momentum $k_\ast$ are amplified in each oscillation,
and the typical number density for one degree of freedom can be evaluated as
\begin{align}
	n_\chi \sim \frac{k_\ast^3}{(2 \pi)^3} 
	\sim \frac{(\lambda \tilde \phi m_\phi)^{3/2}}{8 \pi^3};
	\label{n_chi}
\end{align}
in each oscillation.

After the $\phi$ passes through the origin, the produced particles become
heavy due to the field value of oscillating $\phi$, and their decay rate 
becomes large correspondingly. 
If the decay rate of $\chi$: $\Gamma_\chi$ is sufficiently large,
they eventually decay into the other light degrees of freedom
at $\Gamma_\chi(\phi (t_{\rm dec}))\, t_{\rm dec} \sim 1$
well before the $\phi$ reaches its maximum value~\cite{Felder:1998vq}.
This is the case for $m_\phi \ll \alpha \lambda \tilde \phi$, if we assume that
the typical decay rate of $\chi$ is given by $\Gamma_\chi \sim \alpha m_\chi
\sim \alpha \lambda |\phi (t)|$. 
In this case, a linear potential for $\phi$ generated by the produced particles~\cite{Kofman:1994rk}
is insignificant, since the produced particles decay
when the inflaton field value reaches $\phi \sim [m_\phi \tilde \phi / (\alpha \lambda)]^{1/2} \ll \tilde \phi$.

In addition, if the inflaton couples to the bosonic fields $\tilde \chi$,
this condition guarantees the absence of violent parametric resonant phenomena.
Unless the coupled bosonic particles decay before the inflaton moves back to the origin,
the production rate of these bosonic particles is enhanced due to the induced emission effect.
Thus, their number density grows exponentially whenever the inflaton passes through the origin,
and as a result,
the system may enter the so-called turbulent regime
as shown in Refs.~\cite{Micha:2002ey} with classical lattice simulations 
and recently  in Refs.~\cite{Berges:2008wm} with Kadanoff-Baym eqs.

Throughout this paper, we assume that the condition $m_\phi \ll \alpha \lambda \tilde \phi$ holds,
and hence such effects can be neglected.\footnote{
	The number density of particles produced by the decay of $\chi$ fields directly coupled to the inflaton
	cannot become so large since otherwise the non-perturbative production becomes inactive
	due to the large screening mass: $m_{\rm s} \gtrsim k_\ast$ [cf. Eq.~\eqref{eq:np_cond}].
}
The energy density converted to the other light
degrees of freedom in one inflaton oscillation can be evaluated as
\begin{align}
	\delta \rho \sim \left. m_\chi n_\chi \right|_{\rm dec} \sim 
	\alpha^{-1/2} (\lambda \tilde \phi m_\phi)^2.
\end{align}

\subsubsection{Thermalization of plasma} 

In order to study the subsequent evolution of the produced plasma and oscillating scalar field $\phi$ at every moment,
it is practically important to know whether or not the produced other light degrees of freedom
can attain thermal equilibrium in a time scale faster than the oscillation time scale
of $\phi$~\cite{Allahverdi:2011aj}. 

At the first passage of $\phi \sim 0$,
the total energy density of light degrees of freedom $\rho_{\rm rad}$ is given by
$\rho_{\rm rad} = \delta \rho$, since there are no particles before this non-perturbative 
particle production.
In this case, as extensively discussed in Ref.~\cite{Kurkela:2011ti},
the thermalization time scale of light degrees is estimated by
\begin{align}
	t_{\rm eq} \sim \left( \alpha^{2} T_{\rm f} \right)^{-1} \sqrt{Q/T_{\rm f}} 
	\sim  \alpha^{-33/16} (\lambda \tilde \phi m_\phi)^{-1/2},
	\label{eq:thrm}
\end{align}
where $T_{\rm f} \sim \rho_{\rm rad}^{1/4}$ and the typical momentum scale $Q$ is given by 
$Q \sim \left. m_\chi \right|_{\rm dec} \sim \alpha^{-1/2} (\lambda \tilde \phi m_\phi)^{1/2}$.
[See Appendix~\ref{sec:thermalization} for more detail.]
Therefore, the produced other light degrees of freedom can be
safely regarded as ``thermal'' plasma as far as the following condition is met:
\begin{align}
	1 \ll \alpha^{33/16} (\lambda \tilde \phi / m_\phi )^{1/2}.
	\label{eq:thrm_cond}
\end{align}
As one can see, this condition is satisfied if the initial amplitude of oscillating
scalar field $\phi_i$ is sufficiently large.

After several oscillations, the energy density of background thermal plasma becomes 
much larger than the one produced via the non-perturbative production in each oscillation, 
{\it i.e.} $\rho_{\rm rad} \gg \delta \rho$.
In this case, the equilibration time is given by the relaxation one,
corresponding to a time scale for a hard particle $Q > T$
to emit its energy away to the thermal plasma of temperature $T$:\footnote{
	Note that this temperature $T$ is not related to the $Q$, contrary to the $T_{\rm f}$.
}
\begin{align}
	t_{\rm rlx} \sim (\alpha^2 T)^{-1} \sqrt{Q/T}.
\end{align}
Hence, the background plasma can remain in thermal equilibrium
if $t_{\rm rlx} \ll m_\phi^{-1}$.

If these conditions are met,
the produced light particles attain thermal equilibrium in each oscillation, and consequently
the screening mass of coupled $\chi$ field can be described by the thermal mass $m_\chi \sim g T$.\footnote{
	Otherwise, the effective mass for $\chi$ cannot be described by a temperature $T$
	and following analyses become more complicated. We do not go into such a case in this paper.
} 
As mentioned above [Eq.~\eqref{eq:np_cond}], the instant preheating stage finishes when
this thermal mass becomes comparable to $k_\ast \sim (\lambda 
\tilde \phi m_\phi)^{1/2}$.

\subsection{Dissipation to thermal plasma} 

As discussed in the previous subsection, the background thermal plasma
is produced via the instant preheating if $\lambda\phi_i > m_\phi$, just after inflation.
The produced thermal plasma can significantly affect subsequent dynamics of oscillating inflaton
field. Aside from the blocking effect on the non-perturbative production
due to the thermal mass, which we discussed in the previous section,
there are basically two effects from thermal environment:
(i) thermal effective potential for the inflaton and (ii) dissipation of the inflaton to thermal plasma.
We are mainly interested in the situation where the scalar field dominates
the Universe, and hence let us discuss the latter effect (ii).\footnote
{
	If the scalar field $\phi$ oscillates dominantly with the thermal potential ({\it e.g.} thermal mass or
	thermal log), its energy density is bounded as $\rho_\phi \lesssim T^4$~\cite{Mukaida:2012qn}.
	Therefore, it is typically less than the energy density of thermal plasma:
	$\rho_{\rm rad} \sim g_\ast T^4 > \rho_\phi$.
}

We will not perform detailed calculations of the dissipation coefficient
and will not show all the list of dissipation coefficients in various regimes in this section.
Instead, let us explain its typical behavior and intuitive physical interpretation
relevant to our following discussion.
We refer to Refs.~\cite{Drewes:2010pf,Mukaida:2012qn,Berera:1995ie+X,BasteroGil:2010pb} 
for basic formalism to calculate them. 
The complete list of dissipation coefficient in various regimes
is summarized in Appendix~\ref{sec:diss_coef}.

The effect of thermal plasma becomes significant in the case where 
the typical time scale of oscillation is much slower than that of interaction
in thermal plasma, {\it i.e.} $m_\phi \ll \alpha T$.
In this case, the oscillating scalar field $\phi$ cannot decay into light 
degrees of freedom, since these would-be decay products acquire thermal masses
which are larger than the $\phi$ mass.
However, the oscillating scalar can dissipate into thermal plasma through
multiple scattering by light particles in thermal plasma,
or more precisely through thermal width of each quasi-particle excitation.

The dissipation coefficient depends on the value of $\phi$,
and its dependence can be divided into two regimes: (i) small field value regime 
$\lambda \phi \ll T$ and (ii) large field value regime $\lambda \phi \gg T$.
In the case (i), the coupled $\chi$ particles are relativistic and 
its number density is given by $T^3$. Hence, the oscillating scalar 
dissipates its energy through scatterings involving $\chi$ particles.
On the other hand, in the case (ii), such processes are unlikely to occur
since the $\chi$ particles become very heavy due to the amplitude of $\phi$
and the number density of $\chi$ is exponentially suppressed correspondingly. 
Therefore, the oscillating scalar dissipates its energy mainly by
multiple scattering of gauge bosons through a higher dimensional operator
obtained from integrating out the heavy $\chi$ field.
Consequently, the dissipation coefficient for $m_\phi \ll \alpha T$
can be evaluated as~\cite{Mukaida:2012qn}
\begin{align}
	\Gamma_\phi \sim 
	\begin{cases}
		A_0\,{\rm dim} (r) \lambda^2 \alpha T / (2 \pi^2) &\mbox{for}~\lambda \phi \ll m_{\rm th}^\chi \sim gT \\
		A_0\,{\rm dim} (r) \lambda^4 \phi^2 / (\pi^2 \alpha T) 
		&\mbox{for}~m_{\rm th}^\chi   \sim g T  \ll \lambda\phi \ll T\\
		b \alpha^2 T^3/\phi^2 &\mbox{for}~\lambda \phi \gg T
	\end{cases}
\end{align}
where
\begin{align}
	b := \( \frac{ {\rm T} (r)}{16 \pi^2} \)^2 \frac{(12 \pi)^2}{\ln \alpha^{-1}}.
\end{align}
Here ${\rm dim}(r)$ stands for the dimension of $\chi$'s representation $r$
of gauge group and ${\rm T} (r)$ is the index of $\chi$'s representation $r$
defined by ${\rm T} (r) \delta^{ab}=  {\rm tr} [t^a (r)t^b (r)]$, 
and $A_0$ is a numerical constant, typically $A_0 \sim 1/2$.
For our numerical calculation, we take $\alpha = 0.05$, and then it is given by
$A_0 \simeq 0.3$.
Note that the above dissipative coefficient is calculated in two limits: large and small amplitude,
and hence we have some ambiguities in the intermediate regime.\footnote{
In addition, the small amplitude result computed with one-loop approximation
may change by some factors due to the resummation of infinitely many higher-loop diagrams
as discussed in Ref.~\cite{BasteroGil:2010pb}.
}

In the opposite limit, $m_\phi > T $, 
the dissipation coefficient can be estimated with neglecting
the finite density correction to the dispersion relation of $\chi$.
Therefore, it is simply given by the perturbative decay rate of 
oscillating scalar into light degrees of freedom.
If the amplitude $\tilde \phi$ is much larger than the mass of $\phi$ 
({\it i.e.} $\lambda \tilde \phi \gg m_\phi$),
the oscillating scalar $\phi$ loses its energy mainly via the non-perturbative 
particle production as discussed in the previous section.
Hence, practically, the pertrurbative decay becomes important at $\lambda \tilde \phi \ll m_\phi$
and it is given by 
\begin{align}
	\Gamma_\phi = {\rm dim} (r) \frac{\lambda^2 m_\phi}{8 \pi}.
\end{align}

In the intermediate region: $\alpha T \lesssim m_\phi \lesssim gT$,
the perturbative decay is kinematically suppressed due to thermal masses of quasi-particles,
and a non-zero dissipation rate comes from their thermal widths~\cite{Yokoyama:2005dv,Drewes:2010pf}. 
As a result, the dissipation coefficient can be approximately expressed as
\begin{align}
	\Gamma_\phi \simeq {\rm dim} (r)
	 \begin{cases}
	 	\cfrac{\lambda^2 m_\phi}{8 \pi} \sqrt{1 -  4 \cfrac{m^\chi_{\rm th}{^2}}{m_\phi^2}}
	 	\left[ 1 - 2 f_{\rm FD} (m_\phi / 2) \right] & {\rm for}~~~m_\phi > 2 m_{\rm th}^\chi \\
	 	\cfrac{\lambda^2 \alpha T}{2 \pi^2} \(
			A_0 + A_1 \left[ \cfrac{m_\phi}{\alpha T} \right]^2 + \cdots
		\) &{\rm for}~~~2 m_{\rm th}^\chi < m_\phi,
	 \end{cases}
\end{align}
where $f_{\rm FD}$ denotes the Fermi-Dirac distribution.
$A_0$ and $A_1$ are numerical constants,
and in our numerical computation with $\alpha = 0.05$,
they are given by $A_0 \simeq 0.3$ and $A_1 \simeq 2\times 10^{-4}$ respectively.
Note that this result is applicable to all the $m_\phi$ region
in a small amplitude regime: $\lambda \tilde \phi < m_{\rm th}^\chi \sim gT$.

\section{Reheating after inflation} 
\label{sec:reh_after_inf}

In the last two subsections, we have introduced basic ingredients to study
the dynamics of oscillating inflaton.
In this section, let us study the dynamics of oscillating inflaton and numerically
evaluate the reheating temperature with some examples.

\subsection{Effective dissipation rate of the inflaton} 

The equation of motion of the inflaton is given by
\begin{align}
	\ddot \phi + ( 3 H + \Gamma_\phi) \dot \phi + m_\phi^2 \phi = 0,
\end{align}
where $\Gamma_\phi$ is the $\phi$-dependent dissipation coefficient and $H$ is the Hubble parameter.
In order to study the dynamics of oscillating scalar field,
it is convenient to consider quantities averaged over a time interval that is  longer than
the oscillation time scale but shorter than the Hubble and dissipation time scale.
Taking this time average, one finds 
\begin{align}
	&\dot \rho_\phi + 3 H \rho_\phi =  - \Gamma_\phi^{\rm eff} \rho_\phi, \label{eq:diff_inf} \\
	&\dot \rho_{\rm rad} + 4 H \rho_{\rm rad} = \Gamma_\phi^{\rm eff} \rho_\phi, \label{eq:diff_rad}\\
	&3 M_{\rm pl}^2 H^2 = \rho_\phi + \rho_{\rm rad} \label{eq:diff_const},
\end{align}
where $M_{\rm pl}$ is the reduced Planck mass, 
$\rho_{\rm rad}$ stands for the energy density of radiation,
the energy density of inflaton is given by
$\rho_\phi := \overline{\dot \phi^2/2 + m_\phi^2 \phi^2/2}$ and 
the effective dissipation rate is defined as
$\Gamma^{\rm eff}_\phi := \overline{\Gamma_\phi \dot{\phi^2}}/ \overline{\dot {\phi}^2}$.
Here the time-averaged quantity is represented by $\overline{\cdots}$.
Since $\Gamma_\phi$ depends on $\phi$, $\Gamma_\phi^{\rm eff}$
is different from $\Gamma_\phi$ in general.
Let us summarize the effective dissipation rate $\Gamma_\phi^{\rm eff}$ relevant to
our following discussion.
The complete list is shown in Appendix~\ref{sec:diss_coef}.

The effective dissipation coefficient is independent of the amplitude $\tilde \phi$
in the the small amplitude regime
$(\lambda \tilde \phi \lesssim m_{\rm th}^\chi \sim gT)$ [See Eq.~\eqref{eq:diss_small}]:
\begin{align}
	\Gamma^{\rm eff}_\phi
	\simeq {\rm dim} (r)
	\begin{cases}
		\cfrac{\lambda^2 m_\phi}{8 \pi} \sqrt{1 - 4 \cfrac{m_{\rm th}^\chi{^2}}{m_\phi^2}}
		\left[ 1 - 2 f_{\rm FD} (m_\phi /2) \right] 
		&\mbox{for}~~2m_{\rm th}^\chi < m_\phi\\[20pt]
		\cfrac{\lambda^2 \alpha T}{2 \pi^2}
		\( A_0 + A_1 \left[ \cfrac{m_\phi}{\alpha T} \right]^2 \)
		&\mbox{for}~~m_\phi < 2 m_{\rm th}^\chi.
	\end{cases}
\end{align}
For our numerical computation, we take $\alpha = 0.05$,
and then the numerical constants $A_0$ and $A_1$ are given by
$A_0 \simeq 0.3$ and $A_1 \simeq 2\times 10^{-4}$ respectively.
The effective dissipation rate for non-perturbative production is also
independent of $\tilde \phi$ and it is given by
\begin{align}
	\left. \Gamma_\phi^{\rm eff}\right|_{\rm NP}
	=
	{\rm dim} (r)
	\cfrac{\lambda^2 m_\phi} {\pi^4 \sqrt{\alpha}}
	~~\mbox{for}~~\lambda \tilde \phi \gg {\rm max} \left[ m_\phi, \cfrac{g^2 T^2}{m_\phi} \right].
\end{align}

On the other hand, if the amplitude $\tilde \phi$ is larger than $m_{\rm th}^\chi$,
then the effective dissipation coefficient depends on $\tilde \phi$.
In particular, the effective dissipation coefficient for the regime $m_\phi < \alpha T$
is important in the following discussion.
In this case, the effective dissipation coefficient can be approximated as [Eq.~\eqref{eq:diss_large}]
\begin{align}
	\Gamma_\phi^{\rm eff}
	\simeq C\,
	\frac{4}{3 \pi^3} \tilde A_0\, {\rm dim} (r)
	\frac{\lambda}{\alpha} \frac{T^2}{\tilde \phi}
	~~\mbox{for}~~\lambda \tilde \phi \gg T.
	\label{eq:diss_th}
\end{align}
Here we have explicitly included an uncertainty in a numerical constant $C$ that
is caused by the extrapolation in the intermediate regime as we already mentioned,
and it is taken to be unity: $C=1$ for our numerical computation.
Practically, the dissipation coefficient for $m_{\rm th }^\chi < \lambda \tilde \phi < T$ is not 
important, since this term $\Gamma^{\rm eff}_\phi \propto \tilde \phi^2$ [See Eq.~\eqref{diss_mid_2}]
cannot complete the reheating of the Universe because it decreases faster than the Hubble parameter.
[See also Fig.~\ref{fig:evolution} and footnote~\ref{ft:def_TR}.]

\subsection{Evolution of inflaton/plasma system and reheating} 

\begin{figure}
\begin{center}
\vskip -1.cm
\includegraphics[scale=1.2]{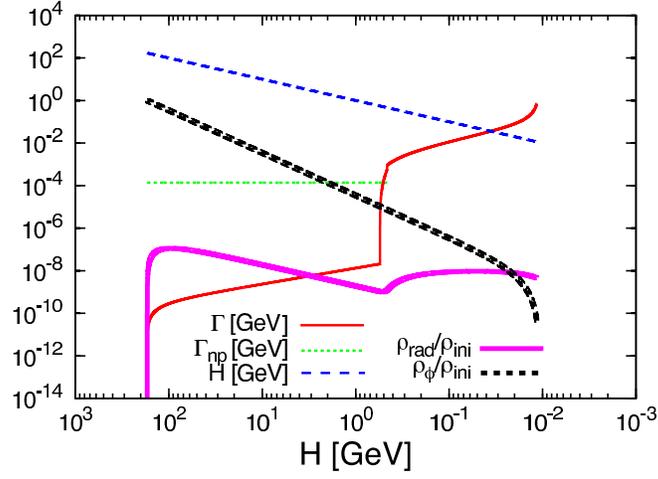}
\vskip 0.3cm
\includegraphics[scale=1.2]{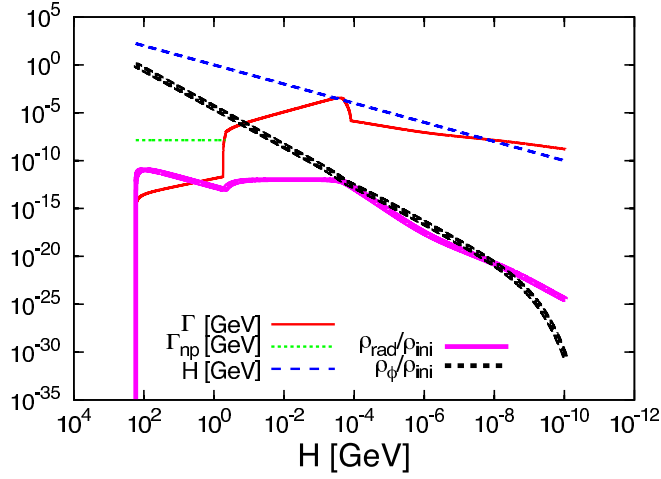}
\vskip 0.3cm
\includegraphics[scale=1.2]{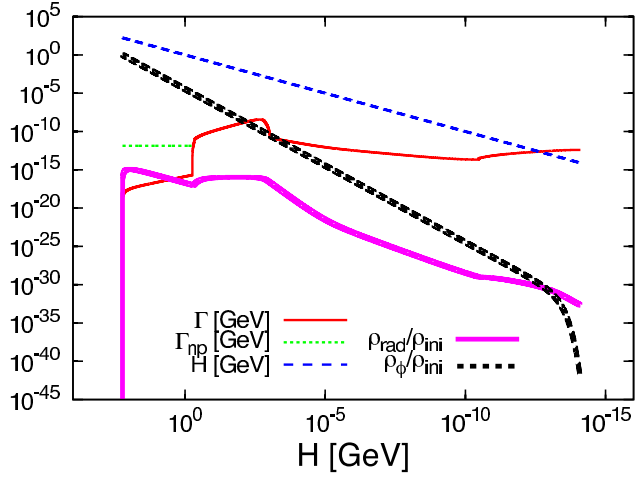}
\vskip -0.1cm
\caption{\small The evolution of various quantities as a function of Hubble scale $H$:
the effective dissipation rate except for non-perturbative particle production $\Gamma$ ({\it \color{red} red thin solid}),
one for non-perturbative particle production $\Gamma_{\rm np}$ ({\it \color{green} green thin dotted}),
the energy density of radiation $\rho_{\rm rad}$ ({\it \color{magenta} magenta thick solid})
and inflaton $\rho_\phi$ ({\it black thick dashed}) normalized by an initial energy density $\rho_{\rm ini}$.
{\bf Top}: $(m_\phi, \lambda, \phi_i) = (1\, {\rm TeV}, 10^{-3}, 10^{18}\, {\rm GeV})$,
{\bf Middle}: $(m_\phi, \lambda, \phi_i) = (1\, {\rm TeV}, 10^{-5}, 10^{18}\, {\rm GeV})$,
{\bf Bottom}: $(m_\phi, \lambda, \phi_i) = (1\, {\rm TeV}, 10^{-7}, 10^{18}\, {\rm GeV})$.
}
\label{fig:evolution}
\end{center}
\end{figure}

Now we are in a position to calculate the evolution of inflaton/plasma system after inflation.
We numerically solve the differential Eqs.~\eqref{eq:diff_inf} -- \eqref{eq:diff_const}
to study the dynamics of inflaton/plasma system,
using the effective dissipation rate summarized in App.~\ref{sec:diss}.
To make our discussion concrete, let us assume that $\chi$ is an extra quark like matter
charged under ${\rm SU} (3)$ with the fundamental representation in the following.
Then, we have ${\rm dim} (3) = 3$ and ${\rm T}(3) = 1/2$.
And the gauge coupling constant is assumed to be $\alpha = 0.05$ hereafter.

Fig.~\ref{fig:evolution} shows the evolution of various quantities
as a function of Hubble scale $H$:
the effective dissipation rate $\Gamma^{\rm eff}_\phi$ except for non-perturbative particle
production, that for non-perturbative particle production
$\left. \Gamma^{\rm eff}_\phi \right|_{\rm NP}$,
the energy density of radiation $\rho_{\rm rad}$ and inflaton $\rho_\phi$,
normalized by an initial energy density $\rho_{\rm ini} (=m_\phi^2 \phi_i^2/2)$.

The top panel is computed with $(m_\phi, \lambda, \phi_i) = (1\, {\rm TeV}, 10^{-3},10^{18}\, {\rm GeV})$.
First, the radiation with high temperature ($T \sim 10^8\, {\rm GeV}$) 
is produced via the instant preheating.
The condition for non-perturbative production [Eq.~\eqref{eq:np_cond}] soon saturates,
since the amplitude scales as $\tilde \phi \propto a^{-3/2}$ where $a$ is the scale factor of the Universe
while the temperature scales as $T \propto a^{-3/8}$,
and consequently the non-perturbative production shuts off.
Then, as can be seen from the plateau of $\rho_{\rm rad}$ around 
$H \sim 5 \times 10^{-1}$ -- $10^{-2}\, {\rm GeV}$, the temperature of thermal plasma becomes
nearly constant during the regime where the dominant dissipation rate is given by 
$\Gamma^{\rm eff}_\phi \sim \lambda T^2 /( \alpha \tilde \phi)$.
At this regime, the energy density of radiation behaves as
$\rho_{\rm rad} \sim \Gamma_\phi^{\rm eff} \rho_\phi / H 
\sim M_{\rm pl}^2 T^2  (H / \tilde \phi)$. Therefore, the temperature becomes constant
since the inflaton dominates the Universe at that time.
Finally, the reheating is completed at $\Gamma_\phi^{\rm eff} \sim H$.
In the top panel, the reheating takes place via
$\Gamma_\phi^{\rm eff} \sim \lambda T^2/ (\alpha \tilde \phi)$ 
at $H \sim 2 \times 10^{-2}\, {\rm GeV}$,
and the reheating temperature is $T_{\rm R} \sim 10^8\,{\rm GeV}$.

The middle [bottom] panel is computed with 
$(m_\phi, \lambda, \phi_i) = (1\, {\rm TeV}, 10^{-5},10^{18}\, {\rm GeV})$
[$(m_\phi, \lambda, \phi_i) = (1\, {\rm TeV}, 10^{-7},10^{18}\, {\rm GeV})$].
The subsequent evolution is the same as the top panel case in the both middle and bottom panels.
First, the thermal plasma is produced via the instant preheating,
and the condition for non-perturbative production soon saturates.
Then, the plateau region follows $H \sim 5 \times 10^{-1}$ -- $10^{-4}\, {\rm GeV}$
[$H \sim 5 \times 10^{-1}$ -- $10^{-3}\, {\rm GeV}$].
After that, since $\tilde \phi$ decreases due to the cosmic expansion,
the dominant dissipation rate becomes $\Gamma^{\rm eff}_\phi \sim \lambda^2 \alpha T$.
In the middle panel, the reheating takes place via $\Gamma^{\rm eff}_\phi \sim \lambda^2 \alpha T$
at $H \sim 10^{-8}\,{\rm GeV}$,
and the reheating temperature is $T_{\rm R} \sim 10^{5}\, {\rm GeV}$.
On the other hand, in the bottom panel, the reheating occurs via $\Gamma_\phi^{\rm eff} \sim 
\lambda^2 m_\phi$ at $H \sim 10^{-13}\, {\rm GeV}$,
and its temperature is given by $T_{\rm R} \sim 3 \times 10^{2}\, {\rm GeV}$.\footnote{
	Usually, the reheating temperature $T_{\rm R}$ is defined as the temperature at which the radiation dominated Universe 
	begins and it roughly corresponds to the epoch $H\sim \Gamma_\phi$ as (\ref{TR_wb}).
	In the present situation with thermal dissipation effect,
	this definition is ambiguous because of the peculiar behavior of $\Gamma_\phi^{\rm eff}$.
	As seen in the middle panel of Fig.~\ref{fig:evolution}, $\Gamma_\phi^{\rm eff}$ can once become equal to $H$
	but the relation $\rho_{\rm rad}\sim \rho_\phi$ may hold thereafter without exponential decay of the inflaton for a while.
	This is because the dissipation rate decreases faster than the Hubble parameter
	during the regime: $\Gamma^{\rm eff}_\phi \propto \tilde \phi^2$.
	Therefore, the reheating temperature $T_{\rm R}$ here is defined as the temperature at which the inflaton energy density
	begins to decrease exponentially.
	One should note that, although the parameter $T_{\rm R}$ is a convenient quantity 
	which describes a global picture of the early Universe,
	actual thermal history before the reheating would be significantly different from a conventional one.
	\label{ft:def_TR}
}

\begin{figure}
	\begin{center}
	\subfigure{
		\includegraphics[clip,width=0.48\columnwidth]{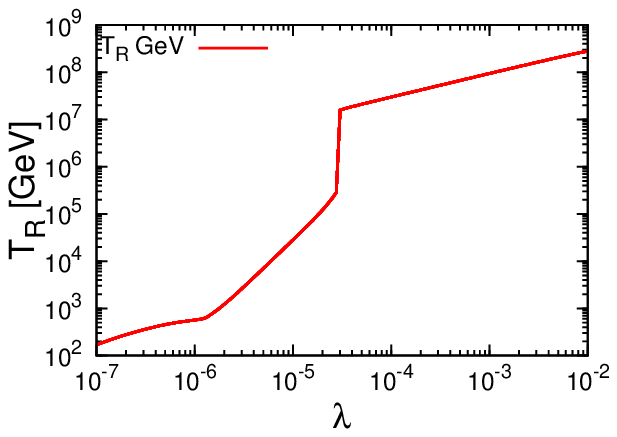}}
		\subfigure{
		\includegraphics[clip,width=0.48\columnwidth]{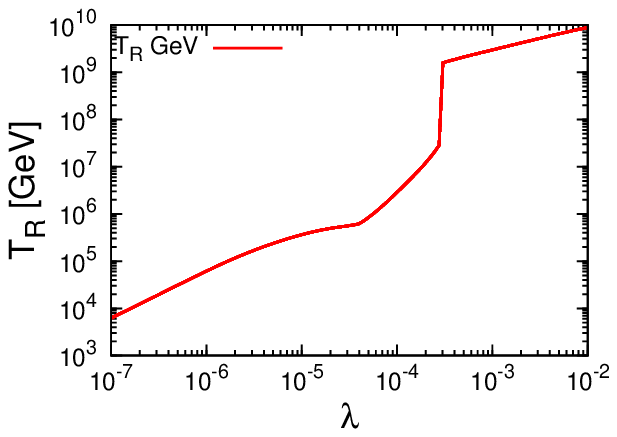}}
		\caption{
		\small The reheating temperature $T_{\rm R}$ as a function of $\lambda$ is shown.
		{\bf Left}: $m_\phi = 1\,{\rm TeV}$ and {\bf Right}: $m_\phi = 10^3\,{\rm TeV}$.}
      		\label{fig:TR}
	\end{center}
\end{figure}

\begin{figure}
\begin{center}
\vskip -1.5cm
\includegraphics[scale=1.5]{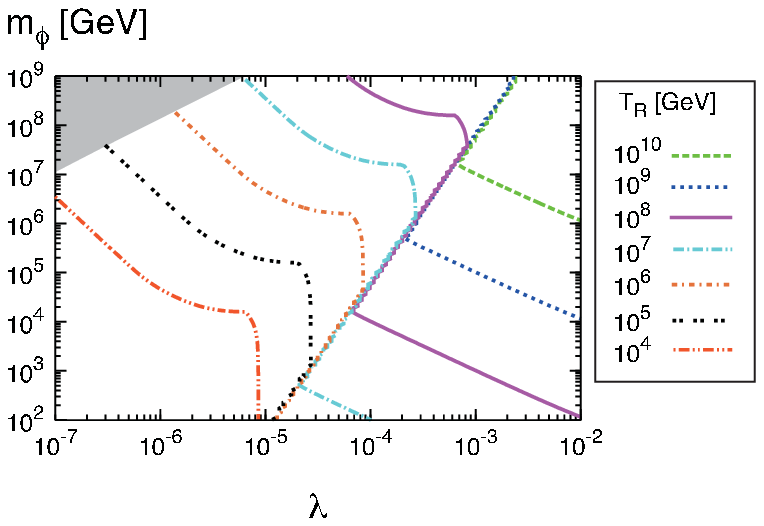}
\vskip 1.0cm
\includegraphics[scale=1.5]{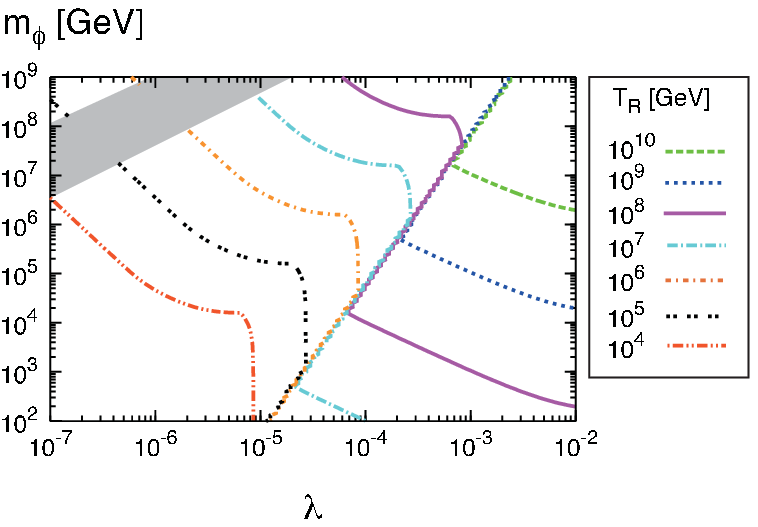}
\caption{\small Contour plot of reheating temperature $T_{\rm R}$
as a function of $\lambda$ and $m_\phi$;
{\bf Top}: $\phi_i = 10^{18}\, {\rm GeV}$ and {\bf Bottom}: $\phi_i =10^{15}\, {\rm GeV}$.
Inside the shaded region, the condition $\lambda \alpha \tilde \phi > m_\phi$
is violated, and this region depends on the initial amplitude $\phi_i$.
At the upper left corner of bottom panel, one can see the region 
where the non-perturbative production is completely absent because $\lambda \phi_i < m_\phi$.}
\label{fig:cntrplot}
\end{center}
\end{figure}

Analytically, the reheating temperature can be roughly estimated as follows in the three cases:
reheating via 
(i) $\Gamma^{\rm eff}_\phi \sim  \lambda T^2 /( \alpha \tilde \phi)$,
(ii) $\Gamma^{\rm eff}_\phi \sim \lambda^2 \alpha T$ 
and (iii) $\Gamma^{\rm eff}_\phi \sim \lambda^2 m_\phi$ :
\begin{align}
	T_{\rm R} \sim
	\begin{cases}
	C^{1/2}
	\(\cfrac{\tilde A_0 {\rm dim} (r)} {g_\ast \alpha}\)^{1/2} \( \lambda M_{\rm pl} m_\phi \)^{1/2}
	~~&\cdots \mbox{(i)} \\
	\( \cfrac{A_0^2 {\rm dim}(r)^2 \alpha^2}{g_\ast} \)^{1/2} \( \lambda^2 M_{\rm pl} \)
	~~&\cdots \mbox{(ii)} \\
	\( \cfrac{{\rm dim} (r)}{g_\ast^{1/2}} \)^{1/2} \( \lambda^2 M_{\rm pl} m_\phi \)^{1/2}
	~~&\cdots \mbox{(iii)}
	\end{cases}
	\label{eq:TR}
\end{align}
Note that the resultant reheating temperature contains the uncertainty $C$ from
Eq.~\eqref{eq:diss_th}. Importantly, the coupling $\lambda$ dependence differs among (i) -- (iii)
and the initial amplitude $\phi_i$ dependence is absent even in the case (i).
These behavior can be seen in Figs.~\ref{fig:TR} and~\ref{fig:cntrplot}.
In Fig.~\ref{fig:TR}, reheating temperature is plotted as a function of $\lambda$ for 
$m_\phi = 1\,{\rm TeV}$ (left) and $m_\phi = 10^3\,{\rm TeV}$ (right) with $\phi_i = 10^{18}\, {\rm GeV}$.
It is seen that in the small $\lambda$ limit, the reheating temperature is determined by the standard perturbative decay scenario
(case (iii)).
As $T_{\rm R}$ increases and approaches to $m_\phi$ for larger $\lambda$, it begins to saturate due to 
the effect of thermal blocking. 
For larger $\lambda$, however, thermal dissipation comes in and again $T_{\rm R}$ increases [case (ii) and (i)].
This figure does not depend on $\phi_i$ for $\phi_i \gtrsim 10^{15}\,{\rm GeV}$.
In Fig.~\ref{fig:cntrplot}, contours of reheating temperature as a function of $\lambda$ and $m_\phi$ are shown.

	As mentioned in footnote~\ref{ft:def_TR},
	the reheating temperature $T_{ \rm R}$ here is defined as the temperature at which
	the inflaton energy density begins to decrease exponentially.
	The sharp discontinuity between two regimes [(i) and (ii)] seen in Fig.~\ref{fig:TR} is related to the definition of reheating 
	temperature $T_R$. The reheating cannot be completed during the regime where
	the effective dissipation rate is given by $\Gamma^{\rm eff}_\phi \propto \tilde \phi^2$,
	with the definition of reheating that we employed.
	This is clearly seen in the middle panel of Fig.~\ref{fig:evolution} $(\lambda=10^{-5})$: $\Gamma$ crosses $H$ two times,
	but at the first crossing the reheating is not completed and the reheating temperature is roughly determined 
	at the second crossing.
	For larger $\lambda$, the situation becomes close to the top panel of Fig.~\ref{fig:evolution} ($\lambda = 10^{-3}$), where
	reheating is completed soon after the first crossing. 
	Thus the reheating temperature jumps somewhere around $10^{-5}< \lambda < 10^{-3}$ (for $m_\phi=1$\,TeV)
	if it is plotted as a function of $\lambda$. This is the reason for the behavior in Fig.~\ref{fig:TR}.

Here we note that, for the purpose of estimating the reheating temperature,
it is not necessary to impose the condition that the light degrees of freedom should
be always kept in thermal equilibrium for every $\phi$'s oscillation.
The only requirement is that the typical interaction time scale of plasma
becomes much faster than the Hubble parameter and the dissipation rate of $\phi$
before the reheating is completed.

\section{Conclusions and Discussion} 
\label{sec:conclusion}

In this paper we have investigated the issue of reheating and thermalization after inflation.
If the inflaton is not heavy enough and its coupling to light species is not so small, 
the standard reheating scenario in which the perturbative decay of the inflaton triggers the reheating may not hold.
In such a case, we need to carefully study the particle production, thermalization,
and the resulting dissipation effect on the inflaton coherent oscillation.
Actually, we found that the dissipation effect in thermal plasma plays a crucial role in the completion of reheating,
and the reheating temperature can be much higher than the inflaton mass.
This is consistent with the statement of Refs.~\cite{Yokoyama:2005dv,Drewes:2010pf},
but our setup and methods are more general and have broad applicability to concrete models.

For example, in the Higgs inflation~\cite{Bezrukov:2007ep,GarciaBellido:2008ab} scenario 
(see Refs.~\cite{Einhorn:2009bh,Nakayama:2010sk} for its realization in NMSSM), 
the inflaton mass around the vacuum is weak scale.
Although $\phi^4$ potential is dominant after inflation and the evolution of inflaton/plasma system would be
different from our model, the final process of reheating cannot be understood without taking the dissipation effect into account,
as studied in this paper.
MSSM inflation~\cite{Allahverdi:2006iq} is another example in which the inflaton is very light.
Non-perturbative particle production is expected at the first several oscillations~\cite{Allahverdi:2011aj},
but the final reheating may be caused by dissipative effects in thermal plasma.
Similar results may hold for alchemical inflation scenario~\cite{Nakayama:2012gh},
in which the inflaton has a mass of soft SUSY breaking scale and oscillates around the origin after inflation.
It will also be useful for reheating after a class of thermal inflation model~\cite{Hindmarsh:2012wh}.
It should also be noticed that the evolution of thermal plasma before the complete reheating in these cases
can significantly differ from a conventional scenario.
Phenomenological consequences, such as relic abundance of heavy particles created in high-temperature plasma
and the efficiency of baryogenesis,
may not be characterized by a single parameter $T_{\rm R}$, but by a detailed thermal history before reheating.
We will further generalize our results and apply to concrete models in a separate paper.

If the inflaton oscillates around a large VEV as in the case of new inflation~\cite{Linde:1981mu},
it is expected that the dissipation effect would be much milder and the conventional reheating scenario
by the perturbative decay would be appropriate for broader parameter spaces.
This might also be true in hybrid inflation~\cite{Linde:1991km,Copeland:1994vg}, 
although it is rather non-trivial because the inflaton oscillates around the origin
while the waterfall field oscillates around a large VEV.
We leave these issues for a future work.

Finally we comment on the fate of particle-like excitations of inflaton.
In the case where thermal dissipation plays a crucial role for the reheating with $T_{\rm R} > m_\phi$,
the inflaton particles are expected to have (nearly) thermal abundance.
When the temperature decreases to $T\sim m_\phi/20$ after the completion of reheating, 
the inflaton freezes out from thermal bath. 
At this stage, the perturbative decay of the inflaton opens since the temperature is lower than the inflaton mass,
and the decay rate is given by $\sim \lambda^2 m_\phi$,
which is much larger than the Hubble rate at this epoch, $H\sim 10^{-2}m_\phi^2/M_P$
unless $\lambda$ is extremely small.
Therefore, the inflaton particles decay into light species and
disappear as soon as they freeze out from thermal bath.

\section*{Acknowledgment}

This work is supported by Grant-in-Aid for Scientific research from
the Ministry of Education, Science, Sports, and Culture (MEXT), Japan,
No.\ 21111006 (K.N.) and No.\ 22244030 (K.N.) and also 
by World Premier International Research Center
Initiative (WPI Initiative), MEXT, Japan. 
The work of K.M. is supported in part by JSPS Research Fellowships
for Young Scientists.

\appendix

\section{Thermalization} 
\label{sec:thermalization}

In this section, let us briefly review the thermalization of weakly coupled plasma for the case of instant preheating,
following Ref.~\cite{Kurkela:2011ti}.

As discussed in Sec.~\ref{sec:inst_prht},
due to the break-down of adiabaticity of the coupled particles $\chi$ around $\phi \sim 0$,
$\chi$ particles are produced non-perturbatively in each oscillation as shown in Eq.~(\ref{n_chi}).
After the passage of origin $\phi \sim 0$, they become heavier and the decay rate becomes larger correspondingly.
Eventually, they decay into the other light degrees of freedom at $\Gamma_\chi(\phi(t_{\rm dec})) t_{\rm dec} \sim 1$
before the $\phi$ comes back to the origin,
if the decay rate of $\chi$, given by $\Gamma_\chi$, is sufficiently large.
Assuming the typical decay rate of $\chi$ as $\Gamma_\chi \sim \theta^2 \alpha m_\chi \sim \theta^2 \alpha \lambda |\phi (t)|$,
one finds that this is the case for $m_\phi \ll \theta^2 \alpha \lambda \tilde \phi$.
Here $\theta$ denotes the mixing between $\chi$ and other light degrees of freedom.
In this case, energy density of the other light degrees of freedom is given by 
\begin{align}
	\delta \rho \sim \left. m_\chi n_\chi \right|_{\rm dec} 
	\sim
	\theta^{-1}
	\alpha^{-1/2} \lambda^2  m_\phi^2 \tilde \phi^2;
\end{align}
in each oscillation.
The typical momentum of decay products is roughly given by
\begin{align}
	Q \sim 
	\left. m_\chi \right|_{\rm dec} 
	\sim 
	\theta^{-1} \alpha^{-1/2} \lambda^{1/2} m_\phi^{1/2} \tilde \phi^{1/2}.
\end{align}
Therefore, the converted energy density can be expressed as
\begin{align}
	\delta\rho \sim \theta^{3} \alpha^{3/2} Q^4.
\end{align}
Hereafter, we assume that the mixing $\theta$ is ${\cal O}(1)$.
In fact, the size of mixing angle is not so important in the following discussion.
[See footnote~\ref{ft:mixing}.]

At the first passage of $\phi \sim 0$, the total energy density of light degrees of freedom
is estimated as $\rho = \delta \rho \sim \alpha^{3/2} Q^4$.
This is the so-called under occupied case~\cite{Kurkela:2011ti}, 
since the momentum distribution $f$ around the typical 
momentum $Q$ can be evaluated as $f(Q) \sim \alpha^{3/2} < 1$.
It is instructive to compare the typical momentum $Q$ with the ``temperature'', defined as
$T_{\rm f} \sim \rho^{1/4}$.
This temperature can be expressed as $T_{\rm f} \sim \alpha^{3/8} Q$,
and hence it is smaller than $Q$.
This means that the typical phase space distribution of produced particles for the first crossing
of $\phi \sim 0$ are concentrated on the UV regime, 
compared to the thermal equilibrium distribution.
As discussed in~\cite{Kurkela:2011ti},
in this UV dominated case, the subsequent thermalization takes place as follows.\footnote{
	In the following, we will only consider gauge bosons since they dominate the equilibration
	because of the induced emission factor.
}\\

\noindent
(i)~Soft particles are radiated from the hard particles, and a new population around 
small momentum is created. They eventually fall into a thermal-like
distribution below a scale $p_{\rm max}$ :
\begin{align}
	f_{\rm soft}(p) \sim T_\ast / p; \ \ \mbox{for}\ p < p_{\rm max}.
\end{align}
\\

\noindent
(ii)~Then, the typical scale $p_{\rm max}$ evolves towards UV regime.
The evolution of $p_{\rm max}$ is dominated by the elastic scattering with the hard particles.
When $p_{\rm max}$ reaches $T_\ast$, the distribution function becomes comparable to order $1$,
and then the soft sector is partially thermalized. This time scale can be evaluated as
\begin{align}
	t \sim \alpha^{-5/2} Q^{-1}.
\end{align}
At this stage, the soft sector dominates the screening effect, the number density and
the elastic scattering. However, the energy density is still dominated 
by the remaining hard particles.
\\

\noindent
(iii)~Finally, the remaining hard particles lose their energies to the soft ``thermal'' sector
by multiple splittings of daughter particles. The system thermalizes
when this process is completed. The time scale can be estimated as
\begin{align}
	t_{\rm eq} \sim \left( \alpha^{2} T_{\rm f} \right)^{-1} \sqrt{Q/T_{\rm f}} 
	\sim  \alpha^{-41/16} Q^{-1}. \label{eq:thrm_app}
\end{align}
Therefore, the equilibration time scale is determined by the time scale for
the hard particle $Q \gg T_{\rm f}$ to lose its energy in the presence of thermal bath with
temperature $T_{\rm f}$.
\\

\noindent
If the equilibration time scale $t_{\rm eq}$, given by Eq.~\eqref{eq:thrm_app}, is much
smaller than the oscillation time scale $m_\phi^{-1}$,
we  can safely assume that the produced light particles have enough time to thermalize.
This condition is given by\footnote{
If we keep the mixing angle $\theta$,  this condition becomes
\begin{align}
	1 \ll \theta^{1/8} \alpha^{33/16} \sqrt{\lambda \tilde\phi / m_\phi}.
\end{align}
\label{ft:mixing}
}
\begin{align}
	1 \ll \alpha^{33/16} \sqrt{ \lambda \tilde\phi / m_\phi }. \label{eq:thrm_cond_app}
\end{align}

\section{Dissipation coefficient}
\label{sec:diss_coef}

In this section, we summarize the dissipation coefficient for the sake of completeness.

\subsection{Definition of effective dissipation coefficient}
\label{sec:}

The equation of motion for scalar field is given by
\begin{align}
	\ddot \phi + (3 H + \Gamma_\phi) \dot \phi + m_\phi^2 \phi = 0
\end{align}
Here $\Gamma_\phi$ is an amplitude dependent dissipation coefficient.
We want to calculate averaged quantities with a time-interval that is
longer than the oscillation period but shorter than the Hubble time scale and dissipation time scale. 
In the following, this time-average is represented by $\overline{\cdots}$.
The energy density of $\phi$ field is defined by
\begin{align}
	\rho_\phi 
	&:= \overline {\frac{1}{2} \dot \phi^2 + \frac{1}{2} m_\phi^2 \phi^2 } \\
	&= \frac{1}{2} m_\phi^2 \tilde \phi^2 .
\end{align}
Here $\tilde \phi$ represents an amplitude of $\phi$.
Using the virial theorem, one can derive the evolution equation for the energy density:
\begin{align}
	\dot \rho_\phi + 3 H \rho_\phi = - \Gamma_\phi^{\rm eff} \dot \rho_\phi
\end{align}
where the effective dissipation coefficient is defined as
\begin{align}
	\Gamma_\phi^{\rm eff} := \frac{\overline{\Gamma_\phi \dot\phi^2} }{\overline{\dot\phi^2}}.
\end{align}
In general, the dissipation coefficient $\Gamma_\phi$ depends on $\phi$,
and hence the effective dissipation coefficient $\Gamma_\phi^{\rm eff}$
has a non-trivial $\tilde \phi$ ($\rho_\phi$) dependence.

\subsection{List of Dissipation Coefficient}
\label{sec:diss}

Let us summarize the effective dissipation coefficient $\Gamma_\phi^{\rm eff}$ 
as a function of $\tilde \phi$ and $m_\phi$.~\footnote{
	In what follows, we consider the case where the interaction time scale in thermal plasma
	is much faster than the dissipation coefficient of inflaton:
	$\alpha T \gg \Gamma_\phi$.
}\\

\noindent
(i) $\boldsymbol {\lambda \tilde \phi \ll m^\chi_{\rm th} \sim gT}$:
In this case, the effective dissipation coefficient is independent of $\tilde \phi$.
Therefore, it is exactly the same as the dissipation coefficient $\Gamma_\phi$:
\begin{align}
	 \Gamma_\phi^{\rm eff} = \Gamma_\phi = {\rm dim} (r)
	\begin{cases}
		\cfrac{\lambda^2 \alpha T}{2 \pi^2} \left[ A_0 + A_1 \(\cfrac{m_\phi}{\alpha T} \)^2 \right]
		&\mbox{for}~m_\phi < 2 m_\chi^{\rm th} (T) \\[20 pt]
		\cfrac{\lambda^2 m_\phi}{8 \pi} \sqrt{1 - 4 \cfrac{m^\chi_{\rm th}{^2}}{m_\phi^2}}
		\left[  1  -  2 f_{\rm FD} (m_\phi/2) \right]
		&\mbox{for}~2m_\chi^{\rm th} (T) < m_\phi.
	\end{cases}
	\label{eq:diss_small}
\end{align}
$A_0$ and $A_1$ are numerical constants, 
and they are given by $A_0 \simeq 0.3$ and $A_1 \simeq 2 \times 10^{-4}$ in our numerical
calculation with $\alpha = 0.05$.
Note that we neglect the hole contribution for simplicity~\cite{Bellac}.\\

\noindent
(ii) $\boldsymbol {\lambda \tilde \phi \gg m^\chi_{\rm th} \sim gT}$ {\bf and} 
$\boldsymbol{m_\phi \ll \alpha T}$:
In this case, 
the dissipation coefficient $\Gamma_\phi$ relevant to the following calculation
is given by~\cite{Mukaida:2012qn}
\begin{align}
	\Gamma_\phi =
	\begin{cases}
		A_0\, {\rm dim} (r) \cfrac{\lambda^2 \alpha T}{2 \pi^2}
		&\mbox{for}~\lambda \phi \ll m_{\rm th}^\chi \\[5 pt]
		A_0\, {\rm dim} (r) \cfrac{\lambda^4 \phi^2}{ \pi^2 \alpha T}
		&\mbox{for}~m_{\rm th}^\chi \ll \lambda \phi \ll T \\[5 pt]
		\cfrac{b \alpha^2 T^3}{\phi^2}
		&\mbox{for}~ T \ll \lambda \phi
	\end{cases}
\end{align}
where
\begin{align}
	b :=
	\( \frac{{\rm T (r)}}{16 \pi^2} \)^2 \frac{(12 \pi)^2}{\ln \alpha^{-1}},
\end{align}
where ${\rm dim} (r)$ is the dimension of $\chi$'s representation $r$ of gauge group
and ${\rm T}(r)$ is the index of $\chi$'s representation $r$ that is defined by
${\rm T} (r) \delta^{ab} = {\rm tr} [t^a(r) t^b(r)]$.
Note that the above dissipation coefficients are computed with two
limits; small and large amplitude.
Hence we have some ambiguities in the intermediate regime.
Using these equations, one can compute the effective dissipation coefficient
and it is given by
\begin{align}
	\Gamma_\phi^{\rm eff} 
	=\ &
		A_0{\rm dim} (r) \cfrac{\lambda^2 \alpha T}{2\pi^2}
		\left[ \cfrac{x}{\pi/2} + \cfrac{\sin 2x}{\pi}
		\right] \\
		&+ \tilde {A}_0 {\rm dim }(r) \cfrac{\lambda^4 \tilde\phi^2}{4 \pi^2 \alpha T}
		\left[
			\cfrac{y'}{\pi/2} - \cfrac{\sin 4y'}{2 \pi} - \cfrac{x'}{\pi/2} + \cfrac{\sin 4x'}{2 \pi}
		\right] \\
		&+ \frac{b \alpha^2 T^3}{\tilde \phi^2}
		\left[
			\frac{4y}{\pi} - 2 + \frac{4}{\pi \tan y}
		\right]
\end{align}
where $x$ and $y$ are determined case by case as follows.
$\tilde A_0$ is a numerical constant and
it is given by $\tilde A_0 \simeq 0.2$ for our numerical computation with $\alpha = 0.05$.\\

\noindent
(ii-i) $\boldsymbol{\lambda \tilde \phi < g^2 T^2 / m_\phi}$:
In this case, the non-perturbative production does not occur, and hence 
$x$, $x'$, $y'$ and $y$ are given by
\begin{align}
	x = x' =&
	\begin{cases}
		\arcsin \cfrac{m^\chi_{\rm th}}{\lambda \tilde \phi}
		&\mbox{for}~m^\chi_{\rm th} < \lambda \tilde\phi \\
		\cfrac{\pi}{2}
		&\mbox{for}~ m^\chi_{\rm th} > \lambda \tilde\phi 
	\end{cases}\\
	y = y' =&
	\begin{cases}
		\arcsin \cfrac{T}{\lambda \tilde \phi}
		&\mbox{for}~T < \lambda \tilde\phi \\
		\cfrac{\pi}{2}
		&\mbox{for}~ T > \lambda \tilde\phi,
	\end{cases}
\end{align}

It is instructive to study the asymptotic behavior of $\Gamma_\phi^{\rm eff}$ in two cases:
(a) $\lambda \tilde \phi \gg T$ and (b) $m_{\rm th}^\chi \ll \lambda \tilde \phi \ll T$.
In the case of (a), $x=x'$  and $y = y'$ are given by
\begin{align}
	x = x' &\simeq \frac{m_{\rm th}^\chi}{\lambda \tilde \phi} \ll 1, \\
	y = y' &\simeq \frac{T}{\lambda \tilde \phi} \ll 1.
\end{align}
Therefore, the effective dissipation coefficient can be approximated by
\begin{align}
	\Gamma_\phi^{\rm eff}
	\simeq
	\frac{4}{3\pi^3}\tilde A_0\, {\rm dim} (r) \frac{\lambda}{\alpha} 
	 \frac{T^2}{\tilde \phi}
	 ~~\mbox{for}~\lambda \tilde \phi \gg T.
	 \label{eq:diss_large}
\end{align}
On the other hand, in the case of (b),
$x=x'$ and $y = y'$ are given by
\begin{align}
	x=x' &\simeq \frac{m_{\rm th}^\chi}{\lambda \tilde \phi} \ll 1 \\
	y = y' & = \frac{\pi}{2}.
\end{align}
Then, the dominant contribution to dissipative coefficient can be expressed as
\begin{align}
	\Gamma_\phi^{\rm eff} 
	\simeq
	\tilde A_0 {\rm dim} (r) \frac{\lambda^4 \tilde \phi^2}{4 \pi^2 \alpha T}. 
	\label{diss_mid_2}
\end{align}
\\

\noindent
(ii-ii) $\boldsymbol{\lambda \tilde \phi > g^2 T^2 / m_\phi}$:
In this case, the non-perturbative production occurs.
Inside the region $|\phi| < (m_\phi \tilde\phi / \lambda)^{1/2} =: \phi_{\rm NP}$,
the adiabaticity is broken down.
Hence, we cannot use the WKB solutions inside this region.
There are some ambiguities to evaluate the region where
the dissipation coefficient is replaced by one caused by
the instant preheating,
but we simply evaluate the threshold value as $\phi_{\rm NP}$.
Then, one can show that the threshold value $\phi_{\rm NP}$ is always greater than
$m_{\rm th}^{\chi}$ from the inequality for non-perturbative production.
Therefore, one finds $x = 0$ and the remaining $x'$, $y'$ and $y$ are given by
\begin{align}
	\( x', y', y \) = &
	\begin{cases}
		\( 0,~0,~\arcsin \cfrac{k_\ast}{\lambda \tilde \phi} \)
		&\mbox{for}~ \phi_{\rm NP} > T/\lambda \\[20pt]
		\( \arcsin \cfrac{k_\ast}{\lambda \tilde \phi},~
		\arcsin \cfrac{T}{\lambda \tilde \phi},~
		\arcsin \cfrac{T}{\lambda \tilde \phi}  \)
		&\mbox{for}~T/\lambda > \phi_{\rm NP} ~(>  m_{\rm th}^\chi/\lambda).
	\end{cases}
\end{align}
\\

\noindent
(iii) $\boldsymbol{\lambda \tilde \phi \gg m_{\rm th}^\chi \sim gT}$ {\bf and} 
$\boldsymbol{m_\phi > T}$:
In this case, the dissipation is dominated by the perturbative decay.
Thus, the effective dissipation coefficient is given by
\begin{align}
	\Gamma_\phi^{\rm eff}
	=
	\begin{cases}
		\cfrac{\lambda^2 m_\phi}{8\pi}
		&\mbox{for}~\lambda \tilde \phi \ll m_\phi \\[5pt]
		\cfrac{c\alpha^2 m_\phi^3}{\tilde \phi^2}
		\left[
			\cfrac{4 y }{\pi} - 2 + \cfrac{4}{\pi \tan y}
		\right]
		&\mbox{for}~
		\lambda \tilde \phi \gg m_\phi
	\end{cases}
\end{align}
where 
\begin{align}
	y = \arcsin \cfrac{k_\ast}{\lambda \tilde \phi}.
\end{align}
Here the coefficient $c$ is given by
\begin{align}
	c = \frac{{\rm dim} ({\rm Ad}) }{4 \pi} \( \frac{{\rm T}(r)}{ 4 \pi} \)^2
\end{align}
Note that if $\lambda \tilde \phi \gg m_\phi$, 
the non-perturbative production occurs inevitably.
\\

\noindent
(iv) $\boldsymbol{\lambda \tilde \phi > m_{\rm th}^\chi \sim gT}$ {\bf and}
$\boldsymbol{\alpha T < m_\phi <T}$:
This is the missed region of our calculation.
We simply extrapolate between (ii) and (iii) as a rough approximation.



\end{document}